\documentclass[a4paper,11pt]{article}
\pdfoutput=1 
\usepackage{jheppub} 
\usepackage{hyperref}
\usepackage{color}
\usepackage{graphicx}
\usepackage{textcomp}
\usepackage{subfigure}
\usepackage{amsmath,amssymb}
\usepackage{enumerate}
\usepackage{multirow}
\newcommand{\mathsym}[1]{{}}

\newcommand{\ba}{\begin{array}}
\newcommand{\ea}{\end{array}}
\newcommand{\be}{\begin{equation}}
\newcommand{\ee}{\end{equation}}
\newcommand{\beqa}{\begin{eqnarray}}
\newcommand{\eeqa}{\end{eqnarray}}
\def\16{{\bf 16}}
\def\45{{\bf 45}}
\def\3211{SU(3)$_{\rm C} \times$SU(2)$_{\rm L} \times$U(1)$_{\rm B-L} \times$U(1)$_{\rm 3R}$}
\def\u1u1{U(1)$_{\rm B-L} \times$U(1)$_{\rm 3R}$}
\def\51{SU(5)$\times$U(1)$_{\rm X}$ }

\title{A realistic pattern of fermion masses from a five-dimensional SO(10) model}

\author[a,b]{Ferruccio Feruglio,}
\author[b]{Ketan M. Patel,}
\author[a,b]{Denise Vicino}

\affiliation[a]{Dipartimento  di Fisica e Astronomia `G. Galilei', Universit\`a di Padova, Via
Marzolo 8, I-35131 Padua, Italy}
\affiliation[b]{Istituto Nazionale Fisica Nucleare, Sezione di Padova, I-35131 Padova,
Italy.}

\emailAdd{feruglio@pd.infn.it}
\emailAdd{ketan.patel@pd.infn.it}
\emailAdd{denise.vicino@pd.infn.it}

\abstract{
We provide a unified description of fermion masses and mixing angles in the framework of a supersymmetric grand unified SO(10) model with anarchic Yukawa couplings of order unity. The space-time is five dimensional and the extra flat spatial dimension is compactified on the orbifold $S^1/(Z_2 \times Z_2')$, leading to Pati-Salam gauge symmetry on the boundary where Yukawa interactions are localised. The gauge symmetry breaking is completed by means of a rather economic scalar sector, avoiding the doublet-triplet splitting problem. The matter fields live in the bulk and their massless modes get exponential profiles, which naturally explain the mass hierarchy of the different fermion generations. Quarks and leptons properties are naturally reproduced by a mechanism, first proposed by Kitano and Li, that lifts the SO(10) degeneracy of bulk masses in terms of a single parameter. The model provides a realistic pattern of fermion masses and mixing angles for large values of $\tan\beta$. It favours normally ordered neutrino mass spectrum with the lightest neutrino mass below $0.01$ eV and no preference for leptonic CP violating phases. The right handed neutrino mass spectrum is very hierarchical and does not allow for thermal leptogenesis. We analyse several variants of the basic framework and find that the results concerning the fermion spectrum are remarkably stable.}

\begin{document} 
\maketitle
\flushbottom

\section{Introduction}
\label{introduction}
The diversity of elementary particles and of their fundamental interactions observed at the energies probed in various experiments so far finds an elegant description in Grand Unified Theories (GUT). 
The unification of strong and electroweak interactions in GUT also leads to the unification of fundamental fermions. Such a unification can be partial or complete depending on the choice of unified gauge symmetry. As it is well known, one of the most attractive choice is the GUT based on the SO(10) group \cite{Fritzsch:1974nn}. All the Standard Model (SM) fermions of a given generation can be accommodated in the \16 dimensional spinorial representation of SO(10), together with an additional fermion singlet under the SM gauge group. This new fermion can be identified as right handed (RH) neutrino, a partner of the weakly charged neutrinos in the seesaw mechanism of type I \cite{Minkowski:1977sc, Yanagida:1979as, Glashow:1979nm, Mohapatra:1979ia, GellMann:1980vs}. 

Like most of the interesting proposals of physics beyond the SM, GUT also suffer from drawbacks. The most serious of them is perhaps the fact that the GUT do not provide a unique way to get the observed diversity in low-energy physics from the unity imposed at high energy. In general the unified gauge symmetry can be broken down to the gauge group of the SM in several different ways.
If a spontaneous breaking is realised, this requires the presence of scalar fields in large representations of the gauge group,  allowing arbitrariness in the construction and also leading to problems like doublet-triplet (DT) splitting \cite{Dimopoulos:1981xm,Masiero:1982fe,Grinstein:1982um} and large enhancement of the unified coupling above the scale of grand unification ($M_{\rm GUT}$).
Another difficulty arises in GUT due to the complete unification of the matter fields. The quarks and leptons exhibit different mixing patterns and it is not obvious how to reproduce this feature in a unified framework. The other aspect of the flavour puzzle is the hierarchy among fermion masses  (see \cite{Feruglio:2015jfa} for recent review on the status of flavour puzzle).  In typical SO(10) GUT \cite{Clark:1982ai,Aulakh:1982sw,Aulakh:2003kg,Bajc:2008dc,Melfo:2010gf,Aulakh:2008sn,
Aulakh:2011at,Grimus:2006bb,Grimus:2008tm,Joshipura:2009tg,Matsuda:2000zp,Matsuda:2001bg,Fukuyama:2002ch,Fukuyama:2004xs,Albright:1998vf,Babu:1998wi,Dermisek:2005ij,Dermisek:2006dc,Barbieri:1996ww,Albright:2000dk,Albright:2001uh,Ji:2005zk,Hagedorn:2008bc,King:2006me,King:2009mk,King:2009tj,Dutta:2009bj,Patel:2010hr,BhupalDev:2011gi,BhupalDev:2012nm,Chen:2003zv,Fukuyama:2012rw} in 4D, the Yukawa couplings can vary in a huge range, ${\cal O}(10^{-6})$ to ${\cal O}(1)$, and no advantage is obtained over the SM in this context, as realized by several dedicated attempts of explaining the fermion mass spectrum in some simple SO(10) models, see for examples \cite{Babu:1992ia,Oda:1998na,Bertolini:2004eq,Babu:2005ia,Bertolini:2005qb,Bertolini:2006pe,Grimus:2006rk,Altarelli:2010at,Joshipura:2011nn,Dueck:2013gca,Altarelli:2013aqa}.  Clearly, lots of improvements and efforts are needed to come up with a realistic and natural theory of flavour based on GUT.

Some of the above issues can be addressed by implementing the program of grand unification in higher space-time dimensions \cite{Kawamura:2000ev,Altarelli:2001qj,Hebecker:2002rc,Hebecker:2001wq,Hall:2001pg,Alciati:2005ur,Kitano:2003cn,Mohapatra:2007qf,Fukuyama:2007ph,Dermisek:2001hp,Kim:2002im,Alciati:2006sw,Feruglio:2014jla}. First of all, by adding a new spatial dimension compactified on an orbifold $S^1/Z_2$, we can break the gauge symmetry by selecting appropriate parities of the gauge fields \cite{Kawamura:2000ev}. Only the gauge fields with even parity survive on the 4-dimensional fixed points (or branes) leaving the corresponding gauge symmetry unbroken. In this way, the breaking of SO(10) down to the Pati-Salam (PS) gauge symmetry \cite{Pati:1974yy}, namely SU(4)$_{\rm C} \times$SU(2)$_{\rm L} \times$SU(2)$_{\rm R}$, have been studied in \cite{Dermisek:2001hp,Kim:2002im,Alciati:2006sw}. Once the symmetry is broken through the boundary conditions, one has the freedom to introduce on the branes scalar multiplets transforming only under the unbroken symmetry.
As it was shown in \cite{Kawamura:2000ev}, this offers an elegant solution to the DT splitting problem. A second important aspect concerns the flavour problem of GUT, which  can greatly 
benefit from the presence of an extra compact dimension. In the framework proposed by Kitano and Li in \cite{Kitano:2003cn}, an SO(10) model in five flat space-time dimension (5D) is realised, with the extra dimension compactified on  $S^1/Z_2$. 
The three generations of matter fields are kept in the bulk and their bulk masses create exponential profiles for the corresponding zero-modes. The inter-generational mass hierarchies is explained by ${\cal O}(1)$ fundamental parameters. The difference between the quarks and leptons is reproduced by spontaneous breaking of the SO(10) symmetry into \51 through a bulk scalar multiplet. A complete and predictive model based on this idea has been constructed in \cite{Feruglio:2014jla}, showing that fermion  masses and mixing patterns can be successfully described in terms of fundamental parameters of  ${\cal O}(1)$.

In this paper, we provide a merger of these two basic ideas. We construct a 5D SO(10) model with N=1 supersymmetry (SUSY) in which the extra dimension is compactified on an orbifold $S^1/(Z_2 \times Z_2')$ \cite{Pomarol:1998sd}. An N=1 SUSY in 5D is equivalent to N=2 SUSY in 4D \cite{ArkaniHamed:2001tb}. The reflection under $Z_2$ breaks one of the SUSY while $Z_2'$ is used to break SO(10) down to the PS gauge symmetry. Thus the effective symmetry on one of the two branes is the PS one with N=1 SUSY. The further breaking of PS to the SM gauge symmetry is implemented by introducing appropriate fields on the brane.  Fermions are described by \16 dimensional representations living in the bulk. As a consequence of the breaking of SO(10) down to the PS symmetry the fermion zero modes fall into multiplets of the PS gauge group, namely  $(4,2,1)$ or $(\bar{4},1,2)$, depending on the $Z_2'$ parity assignment, and a doubling of matter fields per each generation is required. This has the advantage of allowing different profiles for the zero modes of  $(4,2,1)$ or $(\bar{4},1,2)$ in each generation. At this stage quark-lepton unification inherited from the PS symmetry still holds, and a new independent source of breaking of the PS symmetry is required. This is obtained by the vacuum expectation value (VEV) of an adjoint scalar multiplet that spontaneously breaks SO(10) into \51 giving rise to a distinct set of zero mode profiles. Such a breaking is flavour blind, introduces only one new parameter and contributes with different weights to lepton and quarks bulk masses.

The model presented here provides a simple and viable alternative to the modified Kitano-Li (KL) model  constructed by us in \cite{Feruglio:2014jla}, based on the framework proposed in \cite{Kitano:2003cn}. In comparison to that, the current model implements in a simpler way the GUT symmetry breaking and requires representations for the scalar fields with smaller dimensionality. The DT splitting problem does not arise since no color triplet is associated with the weak doublets introduced by us. The simplified scalar spectrum on the brane reduces the number of  non-anarchic free parameters in the theory compared to the modified KL model, providing in principle a more predictive framework for the description of the fermion mass spectrum. While the number of independent parameters is still quite large, not allowing for precision tests of the model, we find that all fermion masses and mixing angles can be described with all the fundamental parameters of the theory of ${\cal O}(1)$. A good agreement of the model with the data can only be obtained with large values of $\tan\beta$, where $\tan\beta$ is the ratio of the VEVs of two Higgs doublets used in the minimal supersymmetric standard model (MSSM).  While both the normal and inverted ordering in the light neutrino masses can be obtained, the normal ordering is considerably less fine-tuned in the anarchic Yukawas. We derive predictions for the CP violating phase in the lepton sector, the amplitude of the neutrinoless double beta decay and masses for the right-handed neutrinos.  Within the same basic setup we also study another realisation of the Yukawa interactions, resulting in a model very similar to the modified KL model with an increased set of free parameters. A quantitative comparison of both the alternatives is also given.

The organization of paper is as follows. We describe the model including the dynamics on bulk and on the branes in the next section. We then discuss how the fermion mass relations arise in the model in section \ref{fermionmasses}. A qualitative comparison between the alternative models is given in this section. In section \ref{analysis}, we provide a detailed numerical analysis of the various options and discuss the results and predictions for the different observables. The study is finally concluded in section \ref{conclusion}.

\section{An SO(10) model in five dimensions}
\label{model}
The model is based on a supersymmetric SO(10) grand unified theory in five space-time dimensions \cite{Dermisek:2001hp,Kim:2002im,Alciati:2006sw}. The extra spatial dimension is compactified on an orbifold $S^1/(Z_2 \times Z_2')$ where $S^1$ represents a circle of radius $R$. A periodic coordinate $y$ parametrizes the circle and the action of the parity $Z_2$ $(Z_2')$ is defined by $y \to -y$ $(y' \to -y')$, where $y'\equiv y- \pi R/2$. Points of the circle related by either $Z_2$
or $Z_2'$ are identified. The interval between the two fixed points $y=0$ and $y=\pi R/2$ can be considered as the fundamental region.
The other fixed points $y= \pi R$ and $y=-\pi R/2$ are identified with the points $y=0$ and $y=\pi R/2$, respectively. 
A generic bulk field $\phi(x,y)$ can be categorized by its transformation properties under $Z_2 \times Z_2'$. Denoting by $P$ and $P'$ the parities under $Z_2$ and $Z_2'$ respectively, a field $\phi_{P,P'}(x,y)$ with given parities $(P,P')$  can be expanded in terms of Fourier series as follows \cite{Alciati:2006sw}:
\beqa \label{generic-expansion}
\phi_{++}(x,y) &=& \sqrt{\frac{1}{2\pi R}} \phi^0_{++}(x) + \sqrt{\frac{1}{\pi R}} \sum_{n=1}^{\infty}\phi^{2n}_{++}(x) \cos\left(\frac{2ny}{R}\right)~, \nonumber \\
\phi_{+-}(x,y) &=&  \sqrt{\frac{1}{\pi R}} \sum_{n=0}^{\infty}\phi^{2n+1}_{+-}(x) \cos\left(\frac{(2n+1)y}{R}\right)~,  \nonumber \\
\phi_{-+}(x,y) &=&  \sqrt{\frac{1}{\pi R}} \sum_{n=0}^{\infty}\phi^{2n+1}_{-+}(x) \sin\left(\frac{(2n+1)y}{R}\right)~,  \nonumber \\
\phi_{--}(x,y) &=&  \sqrt{\frac{1}{\pi R}} \sum_{n=0}^{\infty}\phi^{2n+2}_{--}(x) \sin\left(\frac{(2n+2)y}{R}\right)~.
\eeqa
Here $n = 0,1,2,...$ denotes the different 4D Kaluza-Klein (KK) modes of a given bulk field. In the free theory, upon the compactification, a 4D component $\phi^k(x)$ acquires a mass $k/R$, an integer multiple of the compactification scale $1/R$. Only the field with $(P,P')=(+,+)$ contains a massless mode and it is non-vanishing on both the branes. The field $\phi_{+-}$ ($\phi_{-+}$) vanishes on the $y=\pi R/2$ ($y= 0$) brane, while $\phi_{--}$  vanishes on both the branes.

The theory possesses N=1 SUSY in 5D which corresponds to N=2 SUSY in 4D \cite{ArkaniHamed:2001tb}. We utilize the $Z_2$ symmetry to break N=2 SUSY down to the N=1 SUSY in 4D \cite{Pomarol:1998sd}. In our set-up, the matter and gauge fields propagate in the bulk. We introduce a 16-dimensional hypermultiplet $\16_{\cal H}$ for each SM generation of fermions and 45-dimensional vector-multiplet $\45_{\cal V}$ under N=1 SUSY in 5D. In 4D, these correspond to a pair of  N=1 chiral multiplets for $\16_{\cal H} \equiv (\16, \16^c)$, and a vector and chiral multiplets for $\45_{\cal V} \equiv (\45_V, \45_\Phi)$. 
The breaking of  N=2 SUSY down to the N=1 SUSY in 4D is achieved by assigning even $Z_2$ parity to the $\16$ and $\45_V$ multiplets and odd $Z_2$ parity to their superpartners $\16^c$ and $\45_\Phi$.

The  $Z_2'$ symmetry is used to break the SO(10) gauge symmetry down to the PS symmetry \cite{Dermisek:2001hp,Kim:2002im,Alciati:2006sw}. The PS gauge symmetry is isomorphic to SO(6)$\times$SO(4) and hence the parity assignments with respect to $P'$ should  be appropriately chosen such that the generators of SO(6)$\times$SO(4) remain unbroken. Under  SO(6)$\times$SO(4), the two index antisymmetric SO(10) representation $\45$ decomposes as $(15,1) + (1,6) + (6,4)$. 
The first two submultiplets are taken even and the last one is chosen odd under $Z_2'$. This assignment breaks SO(10) down to the PS group and set to zero all the gauge fields, 
other than those of the PS group, on the $y= \pi R/2$ brane. The gauge interactions on this brane respects only the PS gauge symmetry. On the $y=0$ brane, the full $\45_V$ exists but only the PS gauge fields have massless modes. For these reasons, we call the $y=\pi R/2$ brane ``a PS brane'' while the $y=0$ brane ``an SO(10) brane''. 

Once the $P'$ assignments for the gauge fields are chosen as above, the ones for the matter submultiplets follow from the invariance of the gauge interactions. Under the PS symmetry, the SO(10) $\16$-plet decomposes as $(4,2,1) + (\bar{4},1,2)$. It can be seen from the gauge interactions that $(4,2,1)$ and $(\bar{4},1,2)$ must have opposite $P'$ charges. Therefore only one of the two possesses zero modes and is different from zero on the $y=\pi R/2$ brane.  To accommodate zero modes for a full SM fermion generation we have to double the $\16$-plet \cite{Dermisek:2001hp,Kim:2002im,Alciati:2006sw} and assign mutually opposite $P'$ charges for the PS submultiplets. Therefore, we introduce $\16'_{\cal H}$ per each generation in the bulk with $P$ ($P'$) equal (opposite) to that of the $\16_{\cal H}$.  Notice that this doubling destroys the full quark-lepton unification achieved with only one copy of $\16$-plet per generation. We summarize the $P$ and $P'$ assignment  of all the bulk fields in Table \ref{tab1}.
\begin{table}
\begin{center}
\begin{tabular}{cccc } 
\hline
\hline
 ~~5D N=1~~ & ~~4D N=1~~ & ~~4D N=1 in PS~~ & ~~$(P,P')$~~ \\
\hline
 \multirow{4}{0em}{$\45_{\cal V}$} &\multirow{2}{0em}{$\45_V$} & $(15,1,1)+(1,3,1)+(1,1,3)$ & $(+,+)$  \\ 
& & $(6,2,2)$ & $(+,-)$\\ 
&\multirow{2}{0em}{$\45_\Phi$} & $(15,1,1)+(1,3,1)+(1,1,3)$ & $(-,-)$  \\ 
& & $(6,2,2)$ & $(-,+)$\\ 
\hline
\multirow{4}{0em}{$\16_{\cal H}$} &\multirow{2}{0em}{$\16$} & $(4,2,1)$ & $(+,+)$  \\ 
& & $(\bar{4},1,2)$ & $(+,-)$\\ 
&\multirow{2}{0em}{$\16^c$} & $(4,1,2)$ & $(-,+)$  \\ 
& & $(\bar{4},2,1)$ & $(-,-)$\\ 
\hline
\multirow{4}{0em}{$\16'_{\cal H}$} &\multirow{2}{0em}{$\16'$} & $(4,2,1)$ & $(+,-)$  \\ 
& & $(\bar{4},1,2)$ & $(+,+)$\\ 
&\multirow{2}{0em}{$\16'^c$} & $(4,1,2)$ & $(-,-)$  \\ 
& & $(\bar{4},2,1)$ & $(-,+)$\\ 
\hline
\hline
\end{tabular}
\end{center}
\caption{The parities $P$ and $P'$ of different SO(10) multiplets and their Pati-Salam submultiplets.}
\label{tab1}
\end{table}

We now discuss the symmetry breaking pattern in the model. The SO(10) symmetry is broken down to the PS gauge symmetry on the branes by the action of $Z_2'$. We use the mechanism originally proposed by Kitano-Li in \cite{Kitano:2003cn} to break the PS symmetry down to the \3211 group. This can be achieved if an SU(5) singlet belonging to $\45_\Phi$ develops a vacuum expectation value (VEV) which  breaks SO(10) into \51 in the bulk. The residual symmetry on the branes is \3211 which in turn has to be broken into the SM gauge symmetry by introducing appropriate 4D fields on the brane of interest. We will discuss the brane sector and the breaking of \u1u1 down to U(1)$_{\rm Y}$  later in this section. Let's first discuss in details the dynamics in the bulk.

\subsection{The bulk}
The N=1 SUSY in 5D allows only gauge interactions in the bulk \cite{Pomarol:1998sd}. The $\45_\Phi$ interacts with the chiral multiplets $\16$, $\16'$, $\16^c$ and $\16'^c$ through gauge interactions. The superpotential in the bulk is:
\be \label{suppot-bulk}
{\cal W}_{\rm bulk} = \16^c_i \left[ \hat{m}_i+\partial_y - \sqrt{2} g_5 \,\45_\Phi \right] \16_i + \16'^c_i \left[\hat{m}'_i+\partial_y - \sqrt{2} g_5 \, \45_\Phi \right] \16'_i  ~.
\ee
Here $i=1,2,3$ denotes three generations of matter. The bulk masses can be chosen real and diagonal without loosing generality and are parametrized by $\hat{m}_i$ and $\hat{m}'_i$. The invariance of ${\cal W}_{\rm bulk}$ under $Z_2\times Z'_2$ makes the bulk masses odd under both the parities and they can be expressed as $\hat{m} = m~sgn(y)$ and $\hat{m}' = m'~sgn(y)$, where $m$ and $m'$ are real constants and $sgn(y)$ has period $\pi R$. Performing a KK expansion for the matter fields, namely $\16(x,y) = \sum_n \16_n(x) f_n(y)$, after the dimensional reduction one gets for the massless modes \cite{Kitano:2003cn} :
\be \label{zeroprofile}
f_0(y) = \sqrt{\frac{2 m}{1-e^{-m\pi R}}}~e^{-my}~~~{\rm for}~0 \le y \le \pi R/2~~~.
\ee
The $f_0(y)$ is appropriately normalized in the interval $[0,\pi R/2]$. Similar expression for the profiles of the $\16'$ zero modes can be obtained by replacing $m$ with $m'$ in Eq. (\ref{zeroprofile}).  The 4D massless mode is localized at $y=0$ ($y=\pi R/2$) brane for positive (negative) value of $m$ and its value is exponentially suppressed on the opposite brane. The exponential behaviour of the zero-mode wave-functions can be used to explain the hierarchies among the fermion generations.

The bulk masses do not distinguish the profiles of quarks and leptons of a given generation residing in the $\16$ or $\16'$ and at this stage the
observed differences in the quarks and lepton masses and mixing patterns cannot be reproduced.
A very crucial correction to this picture can be achieved through the Kitano-Li mechanism \cite{Kitano:2003cn}. The VEV of $\45_\Phi$ along the \51 direction introduces a correction to the bulk masses and distinguishes the profiles of the SU(5) submultiplets. As proposed in \cite{Kitano:2003cn}, this correction, which introduces  a single new parameter, modifies the bulk masses according to
\be \label{bulkmass}
m_i^r = m_i - \sqrt{2} Q_X^r g_5 \langle\45_\Phi\rangle~,
\ee
where $r = (10, \bar{5}, 1)$ represent matter SU(5) representations and $Q_X^r$ are the corresponding U(1)$_{\rm X}$ charges: $Q_X^{10}=-1$, $Q_X^{\bar{5}}=3$ and $Q_X^{1}=-5$. The above modification in the bulk masses was argued to be able to generate viable hierarchies in quarks and leptons and this was demonstrated in a specific model \cite{Feruglio:2014jla} through a detailed numerical analysis. Expressing the dimensionful quantities in units of the cut-off scale of the theory $\Lambda$, we rewrite
\be \label{}
a_i^r \equiv \frac{m_i^r}{\Lambda} = \mu_i - Q_X^r k_X~, \ee
where $\mu_i=m_i / \Lambda$ and $k_X = \sqrt{2} g_5 \langle\45_\Phi\rangle/ \Lambda$. 
As discussed earlier, our  $Z'_2$ parity assignment allows massless modes for $(4,2,1) \in \16$, which contains the SM weak doublets of quarks and leptons $(Q,L)$ and for $(\bar{4},1,2) \in \16'$ containing the weak singlet fields $(u^c, d^c, e^c, N^c)$. The different matter fields within PS multiplets receive appropriate corrections from the VEV of $\45_\Phi$ proportional to their U(1)$_{\rm X}$ charges:
\beqa \label{bulkSM}
a_i^Q &=& \mu_i + k_X~;~~a_i^L = \mu_i -3 k_X~; \nonumber \\
a_i^{u^c} &=& \mu'_i + k_X~;~~a_i^{d^c} = \mu'_i -3 k_X~; \nonumber \\
a_i^{e^c} &=& \mu'_i + k_X~;~~a_i^{N^c} = \mu'_i + 5 k_X~. \eeqa
In conclusion $\mu_i$ and $\mu'_i$ are responsible of splitting the profiles with respect to the PS submultiplets while $k_X$ with respect to SU(5) submultiplets. The zero mode profiles for the various matter fields can be rewritten from Eq. (\ref{zeroprofile}) in terms of the dimensioless quantities as:
\be \label{profiles}
n^\alpha_i(y)  \equiv \sqrt{\Lambda} f^\alpha_{0,i}(y) = \sqrt{\frac{2 a_i^\alpha}{1-e^{-a_i^\alpha c}}}~e^{- a^\alpha_i c \frac{y}{\pi R}}~,\ee
where $\alpha = (Q,u^c,d^c,L,e^c,N^c)$ represents MSSM matter fields while $c=\Lambda \pi R$ is a parameter which depends on the relative separation between the compactification scale and cut-off of the theory.

\subsection{The branes}
The N=1 SUSY in 5D forbids Yukawa interactions in the bulk which can be enabled on the branes by introducing a proper Higgs sector. As discussed earlier, on the $y=\pi R/2$ brane 
only the PS gauge symmetry survives and one can introduce 4D fields filling representations of the PS gauge group.  On the contrary, on the $y=0$ brane full SO(10) multiplets of 4D fields are required. Therefore the PS brane provides a more economical option in terms of the number of 4D fields.
More interestingly, for light particles we can introduce only color singlet and electroweak doublet fields on the PS brane, avoiding the DT splitting problem.  We introduce 4D chiral multiplets $H$, $H'$ transforming as $(1,2,2)$, $\Sigma \sim (\bar{4},1,2)$, $\overline{\Sigma}\sim(4,1,2)$ and $T\sim(1,1,3)$ on the PS brane and $\16_H$, $\overline{\16}_H$ on the SO(10) brane. The superpotential is
\beqa \label{W}
{\cal W} &=&  \delta \left(y-\frac{\pi R}{2} \right)~ \frac{1}{\Lambda} \left[ Y_{ij} \16_i \16'_j H  + Y'_{ij} \16_i \16'_j H' + \frac{1}{2} Y_{R\,ij} \16'_i \16'_j \frac{\overline{\Sigma} \,\overline{\Sigma}}{\Lambda} +...\right] \nonumber \\ 
&+& \delta \left(y-\frac{\pi R}{2} \right) ~ w_\pi(H, H', \Sigma, \overline{\Sigma}, T) + \delta(y) ~ w_0(\16_H, \overline{\16}_H)~, \eeqa
where the first line in ${\cal W}$ corresponds to the Yukawa interactions responsible for the masses of matter fields, while $w_\pi$ and $w_0$ are 
superpotentials for the chiral multiplets when the matter fields are turned off. The $Y$ and $Y'$ are complex $3\times 3$ matrices while $Y_R$ is a complex symmetric matrix. Below we discuss the roles played by each of the brane fields.
\begin{itemize}
\item $\Sigma$, $\overline{\Sigma}$ on $y=\pi R/2$ brane
\hfil\break\noindent
These fields on $y=\pi R/2$ brane play a multiple role. As discussed earlier, SO(10) breaks down to \3211. One can construct two orthogonal linear combinations of the generators of the two U(1)'s which can be identified with the generators of U(1)$_{\rm X}$ and the SM hypercharge U(1)$_{\rm Y}$. In our normalization convention, they read
\beqa \label{u1generators}
Q_X &=& 4 \left( T_{3R} - \frac{3}{2} \frac{B-L}{2} \right)~, \nonumber \\
Q_Y &=&  T_{3R} + \frac{B-L}{2}~. \eeqa
The fields $\Sigma$, $\overline{\Sigma}$ take a VEV along the U(1)$_{\rm Y}$ direction, trigger the breaking of \u1u1 down to U(1)$_{\rm Y}$ and contribute to the mechanism by which D-terms are canceled. The VEV of $\45_\Phi$ in the bulk generates D-terms on the branes \cite{ArkaniHamed:2001tb,Barbieri:2002ic,Kaplan:2001ga} associated to the U(1)$_{\rm X}$  gauge symmetry. To preserve SUSY at high scale these D-terms have to be canceled by appropriate dynamics on the branes. The cancellation of the D-term on the $y=\pi R/2 $ brane can be achieved by the VEVs of $\Sigma$ and $\overline{\Sigma}$ with the condition \cite{ArkaniHamed:2001tb,Barbieri:2002ic}:
\be \label{DtermPS}
D_\pi \equiv 2 \langle \45_\Phi \rangle + g_5 Q^\Sigma_X \left( |\langle \Sigma \rangle |^2 - |\langle \overline{\Sigma} \rangle |^2\right)=0~.\ee
Here $Q_X^{\Sigma} = -5$ is the charge under U(1)$_{\rm X}$ of the component of $\Sigma$ that acquires a VEV. Finally, the VEVs of $\Sigma$ and $\overline{\Sigma}$  generate the masses for the right-handed neutrinos as shown in the first line in Eq. (\ref{W}). 

\item $\16_H$, $\overline{\16}_H$ on $y=0$ brane
\hfil\break\noindent
The role of these fields on the $y=0$ brane is similar to that of $\Sigma$ and $\overline{\Sigma}$ on the other brane. 
The VEV of the singlet under \51 residing in $\16_H$, $\overline{\16}_H$ cancels the D-term on $y=0$ brane if 
\be \label{Dterm0}
D_0 \equiv -2 \langle \45_\Phi \rangle + g_5 Q^1_X \left( |\langle \16_H \rangle |^2 - |\langle \overline{\16}_H \rangle |^2\right)=0~,\ee
where $Q^1_X = -5$ is the U(1)$_{\rm X}$ charge of the SM singlet in $\16_H$.

\item $H$, $H'$, $T$ on $y=\pi R/2$ brane
\hfil\break\noindent
The $H$ and $H'$ are responsible for Dirac type masses of all the fermions. Each of the $H$ and $H'$ contains a pair of Higgs doublets which get mixed through the following terms in $w_\pi$ in Eq. (\ref{W}):
\be \label{Higgs-suppot}
w_\pi = \frac{M_H}{2} H^2 + \frac{M_{H'}}{2} H'^2 + m H H' + \lambda T H H' + T (\lambda_H H^2 + \lambda_{H'} H'^2)+...~~~~
\ee
where dots stand for additional terms involving the $\Sigma$, $\overline{\Sigma}$ fields.
Decomposing $H$ and $H'$ into electroweak doublets, $H=(H_u, H_d)$ and $H'=(H'_u,H'_d)$, one obtains the following mass term
after the electroweak singlet in $T$ acquires a VEV:
\be \label{Higgsmassmatrix}
\left( H_u ~~H'_u \right)~ \mathcal{M}~
\begin{pmatrix}
H_d\\
H_d^\prime
\end{pmatrix}~,
~~~~{\rm with}~~ 
\mathcal{M}=
\begin{pmatrix}
M_H  &  m - \lambda \langle T \rangle \\
m + \lambda \langle T \rangle  & M_{H'} \\
\end{pmatrix}~.\ee
Here $M_{H,H'}$ are redefined including the contributions coming from the VEV of $T$. 
All the mass parameters are assumed to be much heavier than the electroweak scale, possibly close to the GUT scale. 
One can arrange a pair of nearly massless Higgs doublets, by enforcing one eigenvalue of ${\cal M}$ being much smaller than the other. Such a pair would be an admixture of doublets residing in $H$ and $H'$ and can be written as 
\be \label{Higgsmixing} 
h_{u,d} = \cos\theta_{u,d} H_{u,d} + \sin\theta_{u,d} H'_{u,d}\ee
where, in the limit $\det({\cal M})=0$, the mixing angles read
$$\theta_{u,d} = \frac{1}{2} \tan^{-1} \left( \frac{2 M_{H'} (m \mp \lambda \langle T \rangle)}{M_{H'}^2 - (m \mp \lambda \langle T \rangle)^2}\right)~.$$
The other combinations orthogonal to $h_u$ and $h_d$ obtain masses as large as the GUT scale. Below the GUT scale, the model contains only one pair $h_{u,d}$ which plays the role of MSSM Higgs doublets and triggers electroweak symmetry breaking. Clearly, getting $h_{u,d}$ much lighter than the GUT scale requires a fine-tuning of the parameters in (\ref{Higgsmassmatrix}). As we show in the next section, both $H$ and $H'$ with $\theta_u \neq \theta_d$ are needed to generate viable quark mixing angles. Hence a non-vanishing $\langle T \rangle$ is required.  We note that the VEV of $T$ breaks SU(2)$_{\rm R}$ by keeping U(1)$_{\rm 3R}$ unbroken and does not give any additional contribution to the D-terms on the PS brane.
\end{itemize}
The model involves multiple scales of symmetry breaking. 
$$ {\rm SO(10)} \xrightarrow{~1/R~} {\rm PS} \xrightarrow{\langle {\rm \45}_\Phi \rangle,\langle T \rangle} {\text{\3211}} \xrightarrow{\langle \Sigma \rangle, \langle \overline{\Sigma} \rangle} {\rm SM} $$
For simplicity, we take all these scales very close to each other and identify them with the GUT scale $M_{\rm GUT}$. Below the GUT scale the theory looks like the MSSM and we expect standard SUSY gauge coupling unification \cite{Giunti:1991ta,Amaldi:1991cn,Langacker:1991an}. In order to suppress the higher order corrections in Eq. (\ref{W}), we take $c \equiv \Lambda \pi R \approx {\cal O}(100)$ so that the cut-off of the theory, $\Lambda$ can be lifted up to the Planck scale (see \cite{Feruglio:2014jla} for more discussions on the allowed range of the $c$ parameter). The higher order corrections are at the percent level and remain smaller than experimental uncertainty in the fermion mass data we adopt. The theory provides a predictive framework for fermion masses and mixing angles,  to be discussed in details in the following section.

Before ending this section we notice that Yukawa interactions can also be present on the SO(10) brane. A possibility is that all Yukawa interactions are localised at $y=0$.
In this case the dynamics on this brane becomes very similar to the one described in the modified Kitano-Li model discussed by us in \cite{Feruglio:2014jla}. The scalar content on the $y=0$ brane in \cite{Feruglio:2014jla} consists of ${\bf 10}_H$, ${\bf 120}_H$, ${\bf 126}_H$, $\overline{{\bf 126}}_H$ and ${\bf 45}_H$. This combination of fields provides the most economical setup for viable fermion masses and mixing angles, a solution of the DT problem using the missing partner mechanism \cite{Masiero:1982fe,Grinstein:1982um,Babu:2006nf,Babu:2011tw} and a consistent GUT symmetry breaking. All these features are already discussed in details in \cite{Feruglio:2014jla} and we do not repeat them here. In the next sections we will briefly comment on the possibility to adopt the same scalar sector for the $y=0$ brane in the present setup and we will study its potential in explaining the fermion masses and mixings. 

\section{Fermion masses on the branes}
\label{fermionmasses}
The bulk and brane superpotentials in Eqs. (\ref{suppot-bulk}) and (\ref{W}) encode the information about the fermion masses and mixing angles. As discussed earlier, the $Z'_2$ parity and the VEV of \45$_\Phi$ split the zero-mode profiles of various fermions, while the mixing of $H$ and $H'$ leads to the following effective 4D Yukawa couplings:
\be \label{4dyukawa}
{\cal Y}_u= F_Q~Y_u~F_{u^c}~;~~{\cal Y}_d= F_Q~Y_d~F_{d^c}~;~~{\cal Y}_e= F_L~Y_d~F_{e^c}~~{\rm and}~~{\cal Y}_\nu= F_L~Y_u~F_{N^c}~, \ee
where ${\cal Y}_{u,d,e,\nu}$ stand for the 3$\times$3 matrices of dimensionless Yukawa couplings of down-type quarks, up-type quarks, charged leptons and Dirac neutrinos, respectively. The profile matrices are given by
\be
\label{Fmat}
F_\alpha =
\left(
\begin{array}{ccc}
n_1^\alpha(\pi R/2) & 0&0\\
0&n_2^\alpha(\pi R/2)  &0\\
0&0&n_3^\alpha(\pi R/2) 
\end{array}
\right)~~~~{\rm with~}\alpha = (Q,u^c,d^c,L,e^c,N^c)
\ee
where $n_i^\alpha(y)$ are defined in Eq. (\ref{profiles}). The $Y_{u,d}$ arise from the mixing of MSSM-like Higgs doublets in $H$ and $H'$ and, from Eq. (\ref{Higgsmixing}), can be explicitly represented in terms of fundamental Yukawas as follows:
\be 
\label{yud}
Y_{u,d} = \cos\theta_{u,d} Y - \sin\theta_{u,d} Y'~.
\ee
Considering the fact that $(F_{u^c})_{33} \approx {\cal O}(1) \gg (F_{u^c})_{22}, (F_{u^c})_{11} $ and the same for $F_{d^c}$, one obtains ${\cal Y}_{u,d} {\cal Y}^\dagger_{u,d} \approx F_Q Y_{u,d} Y_{u,d}^\dagger F_Q^\dagger$. A common Yukawa $Y_u = Y_d$ leads to an unrealistic scenario of nearly vanishing quark mixing angles. Therefore we require (a) at least two pairs of Higgs doublets allowing for different $Y$ and $Y'$ and (b) unequal mixing $\theta_u \neq \theta_d$ to ensure that $Y_u$ and $Y_d$ are different. The latter condition is satisfied in our model by a SU(2)$_{\rm R}$ triplet field $T$ as shown in Eq. (\ref{Higgsmixing}). After the electroweak symmetry breaking through the VEVs of $h_{u,d}$, one obtains the mass matrices: 
\be \label{massmatrices}
M_{d,e} \equiv v \cos\beta ~{\cal Y}_{d,e}~~~{\rm and}~~~ M_u \equiv v \sin\beta ~{\cal Y}_u~, \ee where $\tan\beta \equiv \langle h_u \rangle / \langle h_d \rangle$ and $v \equiv \sqrt{\langle h_u \rangle ^2 + \langle h_d \rangle ^2} =174$ GeV.

The RH neutrinos receive masses through the U(1)$_{\rm B-L}$ breaking VEVs of $\overline{\Sigma}$ and are given as: 
\be 
\label{MR}
M_R \equiv v_R~F_{N^c}~Y_R~F_{N^c}~, 
\ee
where $v_R \equiv \langle \overline{\Sigma}\rangle^2/\Lambda$ represents the seesaw scale. If the cut-off of the theory is raised to the Planck scale, the seesaw mechanism takes place two order of magnitude below the GUT scale,
the right scale to generate viable neutrino masses. The light neutrinos gain masses through the type-I seesaw mechanism and their mass matrix can be expressed as
\be \label{seesaw}
M_\nu \equiv -  \frac{v^2 \sin^2\beta}{v_R} F_L~(Y_u Y_R^{-1} Y_u^T)~F_L~. \ee
The model contains 24 complex parameters of ${\cal O}(1)$ (9 each in $Y$ and $Y'$ and 6 in $Y_R$) as the fundamental Yukawa couplings. In addition, it has two Higgs mixing angles $\theta_{u,d}$ and 7 bulk mass parameters $\mu_i,~\mu'_i$ and $k_X$.

As it was originally proposed  in \cite{Kitano:2003cn}, the bulk masses in Eq. (\ref{bulkSM}) can generate different hierarchies in $F_Q$ and $F_L$, which in turn explain the observed differences in the quark and lepton mixing patterns and mass hierarchies. The SO(10) breaking by $Z'_2$ distinguishes the profiles of left and right handed fields but it still maintains the quark-lepton unification. A milder hierarchy among neutrino masses and large lepton mixing angles result from the VEV of $\45_\Phi$, which distinguishes profiles of different SU(5) submultiplets within the $\16$ and $\16'$. This model differs from the one presented in \cite{Feruglio:2014jla} in the following ways:
\begin{itemize}
\item In comparison to \cite{Feruglio:2014jla}, the current model has three more bulk masses. This provides more freedom in the profiles of zero-mode fermions. For example, the effective SU(5) symmetry in the profiles is broken once $m_i \neq m'_i$ and, unlike in the previous model, one can distinguish between the masses of down-type quarks and charged leptons even if $Y_d = Y_e$.
\item An important difference with respect to \cite{Feruglio:2014jla} is the simplification of the Higgs sector on the brane. In \cite{Feruglio:2014jla}, consistent fermion masses and a solution of the DT splitting problem through the missing partner mechanism required ${\bf 10}_H+{\bf 120}_H$ Higgs representations, which contain three pairs of MSSM-like Higgs doublets. In the current model, only two pairs are required and this reduces the Higgs mixing parameters from eight to two.
\item The scalars introduced on the PS brane are in representations of smaller dimensionality compared to the brane sector fields in \cite{Feruglio:2014jla}. In particular, realistic Yukawa couplings only require a pair of $(1,2,2)$ fields on the PS brane. The DT splitting
is automatically solved since no colour triplets are present in the relevant Higgs multiplets. However we need to arrange only one pair of light doublets and this requires an appropriate potential with a fine-tuning, as explained in the last section.
\end{itemize}

As recalled at the end of the previous section, all Yukawa couplings can be also localised on the SO(10) brane at $y=0$. We can adopt the same scalar sector as in the model discussed in \cite{Feruglio:2014jla}, remarking however a couple of differences 
with respect to our previous model. There are three more bulk masses in the current setup due to the doubling of matter fields in $\16$ and $\16'$ and the Yukawa matrix $Y_{10}$ ($Y_{120}$) is not symmetric (anti-symmetric) in generation space, 
with several new parameters of ${\cal O}(1)$. Clearly, this model does not provide any improvement in comparison to the old model as far as the field content and dynamics on the brane are concerned. It is however characterized by more parameters, which provide more flexibility in reproducing the correct pattern of fermion masses and mixing angles. We will provide a quantitative analysis of this improvement in the next section.

\section{Numerical analysis and results}
\label{analysis}
We now discuss in detail the viability of the model in explaining the observed data of fermion masses and mixing parameters and analyze its prediction for the observables which have not been measured yet. Our approach is similar to the one followed by us earlier in \cite{Feruglio:2014jla}. We take an idealized set of data for fermion masses and mixing parameters extrapolated at the GUT scale in the MSSM and check the viability of the model in reproducing them. As in \cite{Feruglio:2014jla}, we use the results obtained in \cite{Ross:2007az} for the charged fermion masses and quark mixing parameters. The extrapolation was carried out in the MSSM assuming a SUSY breaking scale of about $500$ GeV, and for different values of $\tan\beta$. We perform the viability analysis for two representative values of $\tan\beta$,  10 and 50.  After our previous analysis, the results of the global fit of neutrino oscillation data have been updated \cite{Gonzalez-Garcia:2014bfa} taking into account the most recent data available till the summer 2014. We take these updated low-energy values of neutrino mass squared differences and lepton mixing angles, neglecting RGE corrections. Such an approximation is valid if neutrino masses are hierarchical \cite{Chankowski:2001mx,Antusch:2003kp,Antusch:2005gp} and indeed this is realized in our model as we will show in this section. Following the widely adopted strategy in this kind of analysis
\cite{Bertolini:2006pe,Grimus:2006rk,Altarelli:2010at,Joshipura:2011nn,Dueck:2013gca,
Altarelli:2013aqa}, the data we use are the result of a specific extrapolation and should be taken as a representative set of GUT scale inputs. The actual data depends on features such as the SUSY breaking scale, SUSY scale threshold corrections, which can be estimated only when the exact mechanism of SUSY breaking is known \cite{Hall:1993gn,Blazek:1995nv,Pierce:1996zz}. Keeping these uncertainties in mind, we believe that if a given model can fit a representative set of data very well, then it will be able to reproduce with a similar accuracy and success the actual data, by slightly varying the underlying parameters. We summarize the various observables and their input values in Table \ref{tab2}.
\begin{table}[t]
\begin{small}
\begin{center}
\begin{tabular}{ccc}
 \hline
 \hline
  Observables ~~~~&~~~~$\tan\beta=10$ ~~~~&~~~~ $\tan\beta=50$~~~~\\
 \hline
$y_t $ ~&~ $  0.48\pm 0.02$ ~&~ $ 0.51\pm 0.03$ \\
$y_b $ ~&~ $  0.051\pm 0.002$ ~&~ $ 0.37\pm 0.02$ \\
$y_{\tau} $ ~&~ $  0.070\pm 0.003$ ~&~ $ 0.51\pm 0.04$ \\
 $m_u/m_c$ ~&~ $  0.0027 \pm 0.0006 $ ~&~ $0.0027\pm 0.0006$ \\
 $m_d/m_s$ ~&~ $0.051 \pm 0.007$ ~&~ $0.051 \pm 0.007$ \\
 $m_e/m_{\mu}$ ~&~ $0.0048 \pm 0.0002$ ~&~ $0.0048 \pm 0.0002$ \\
 $m_c/m_t$ ~&~ $0.0025\pm 0.0002 $ ~&~ $0.0023\pm 0.0002 $ \\
 $m_s/m_b$ ~&~ $0.019 \pm 0.002$ ~&~  $0.016 \pm 0.002$ \\
 $m_\mu/m_\tau$ ~&~ $ 0.059 \pm 0.002$  ~&~ $ 0.050 \pm 0.002$ \\
  \hline
 $|V_{us}|$ & \multicolumn{2}{c}{$0.227 \pm  0.001$}  \\
 $|V_{cb}|$ & \multicolumn{2}{c}{$0.037 \pm 0.001 $}  \\
 $|V_{ub}|$ & \multicolumn{2}{c}{$0.0033 \pm 0.0006 $}  \\
 $J_{CP}$ & \multicolumn{2}{c}{$0.000023 \pm 0.000004 $}  \\
 \hline
 $\Delta_S/10^{-5}$ eV$^2$& \multicolumn{2}{c}{$ 7.50 \pm 0.19$ ~(NO or IO)}  \\
 $\Delta_A/10^{-3}$ eV$^2$ & \multicolumn{2}{c}{$2.457\pm 0.047$~(NO)~~~$2.449\pm 0.048$~(IO)}   \\
 $\sin^2\theta_{12}$ & \multicolumn{2}{c}{$0.304\pm0.013$~(NO or IO)}   \\
 $\sin^2\theta_{23}$ & \multicolumn{2}{c}{$0.452\pm0.052$~(NO)~~~$0.579\pm0.037$~(IO)}  \\
 $\sin^2\theta_{13}$ & \multicolumn{2}{c}{$0.0218\pm0.0010$~(NO)~~~$0.0219\pm0.0011$~(IO)}   \\
\hline
\hline
\end{tabular}
\end{center}
\end{small}
\caption{The GUT scale values of the charged fermion masses and quark mixing parameters from 
\cite{Ross:2007az} and neutrino masses and mixing parameters from an up-to-date global fit analysis \cite{Gonzalez-Garcia:2014bfa}. NO (IO) stands for the normal (inverted) ordering in the neutrino masses.}
\label{tab2}
\end{table}
We employ $\chi^2$ minimization technique to fit the free parameters of the models with the data. See the details in \cite{Feruglio:2014jla} for the definition of the $\chi^2$ function and discussion on the optimization technique. 

\subsection{Results for the PS brane}
We first analyze the Yukawa interactions on the PS brane. The compatibility of the model with anarchic Yukawa structure is tested in two ways. We first fit an idealized data set to the model by minimizing the $\chi^2$ with respect to all the free parameters. The range of ${\cal O}(1)$ Yukawa couplings is restricted to be $|Y_{ij}|,~|Y'_{ij}|,~|{Y_R}_{ij}| \in [0.5,~1.5]$ keeping the phases in the full range $[0,~2\pi]$. The aim of this exercise is to assess whether our model can accommodate the data or not.
We carry out this exercise assuming normal (NO) or inverted ordering (IO) in the light neutrino masses and each of the two cases is analyzed for two values of $\tan\beta$. We get poor fits for small $\tan\beta$ corresponding to minimized $\chi^2$ values $\sim 100$ and $\sim 300$ for NO and IO cases respectively. The results for $\tan\beta=50$ are displayed in Table \ref{tab3} for which we get good fits for both NO and IO cases. 
\begin{table}[!ht]
\begin{small}
\begin{center}
\begin{tabular}{c|cc|cc}
 \hline
 \hline
	& \multicolumn{2}{|c|}{Normal ordering} & \multicolumn{2}{|c}{Inverted ordering} \\
~~~\textbf{Observable}~~~ & ~~~\textbf{Fitted value}~~ & ~~\textbf{Pull}~~~&~~~ \textbf{Fitted
value}~~ &
~~\textbf{Pull}~~~\\
 \hline
$y_t $ & 0.51 & 0 & 0.52 & 0.33 \\
 $y_b$ & 0.37 & 0 & 0.38 & 0.50 \\
 $y_{\tau}$ & 0.51 & 0 & 0.51 & 0 \\
$m_u/m_c$ & 0.0027 & 0 & 0.0028 & 0.17\\
$m_d/m_s$& 0.051 & 0 & 0.052 & 0.14\\
$m_e/m_\mu$ & 0.0048 & 0 & 0.0048 & 0 \\
$m_c/m_t$ & 0.0023 & 0 & 0.0023 & 0 \\
$m_s/m_b$ & 0.016 & 0 & 0.017 & 0.50 \\
$m_\mu/m_\tau$& 0.050 & 0  & 0.050 & 0\\
\hline
$|V_{us}|$& 0.227 & 0 & 0.227 & 0\\
$|V_{cb}|$& 0.037 & 0 & 0.037 & 0 \\
$|V_{ub}|$& 0.0033 & 0  &  0.0030 & -0.50 \\
$J_{CP}$ & 0.000023 & 0 & 0.000023 & 0 \\ 
\hline
 $\Delta_S/\Delta_A$  &  0.0305 & 0 & 0.0305 & 0\\
$\sin^2 \theta _{12}$ & 0.304 & 0  &  0.304 & 0 \\
$\sin^2 \theta _{23}$ & 0.452 & 0 & 0.442 & -0.20  \\
$\sin^2 \theta _{13}$ & 0.0218 & 0 &  0.0218 & -0.10 \\
\hline
\hline
$\chi^2_{\rm min}$ & & $\approx 0$& &  $\approx 0.96$\\
\hline
\hline
           & \multicolumn{2}{|c|}{\textbf{Predicted value}} &\multicolumn{2}{|c}{\textbf{Predicted value}}\\
$m_{\nu_{\rm lightest}}$ [meV] & \multicolumn{2}{|c|}{$3.9$} & \multicolumn{2}{|c}{$10.6$} \\
$|m_{\beta\beta}|$ [meV]& \multicolumn{2}{|c|}{4.96} &\multicolumn{2}{|c}{$48.2$}\\
$\sin \delta^l_{CP}$ &\multicolumn{2}{|c|}{-0.39} &\multicolumn{2}{|c}{-0.89}\\
$M_{N_1}$ [GeV]& \multicolumn{2}{|c|}{$190$} & \multicolumn{2}{|c}{$7.12$}\\
$M_{N_2}$ [GeV]& \multicolumn{2}{|c|}{$ 8.02 \times10^5$} & \multicolumn{2}{|c}{$6.75\times10^5$}\\
$M_{N_3}$ [GeV]& \multicolumn{2}{|c|}{$1.43\times10^{14}$} & \multicolumn{2}{|c}{$1.38\times10^{14}$}\\
$\upsilon_R$ [GeV]& \multicolumn{2}{|c|}{$0.04\times 10^{16}$} & \multicolumn{2}{|c}{$0.056\times 10^{16}$}\\
\hline
\hline
\end{tabular}
\end{center}
\end{small}
\caption{Results from numerical fit corresponding to minimized $\chi^2$ for normal (NO)
and inverted ordering (IO) in neutrino masses. The fit is carried out for the GUT scale
extrapolated data given in Table \ref{tab2} for $\tan\beta=50$. The input parameters are
collected in the Appendix.}
\label{tab3}
\end{table}
As it can be seen, all the data are fitted with negligible deviations from their central values. The model parameters obtained at the minimum of $\chi^2$ are listed in the Appendix. The basic features of the best fit results are similar to the ones obtained in the previous model \cite{Feruglio:2014jla}.  The observed hierarchies of quark and lepton masses requires  $|k_X| \sim |\mu_{2,1}|, |\mu'_{2,1}| \ll |\mu_3|, |\mu'_3|$. This in turn enforces a common bulk mass for quarks and leptons of the third generation and leads to approximate Yukawa unification $y_t \sim y_b \sim y_\tau$, which prefers large $\tan\beta$ \cite{Ananthanarayan:1991xp,Ananthanarayan:1992cd}. 

\begin{figure}[!ht]
\centering
\subfigure{\includegraphics[width=0.6\textwidth]{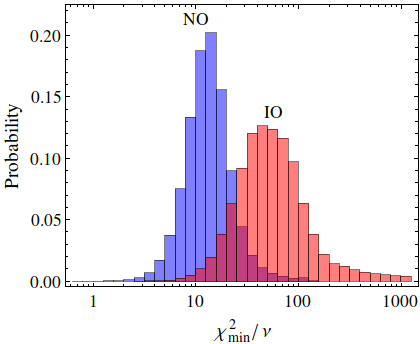}}
\caption{The probability distributions of minimized $\chi^2/\nu$ for NO (blue) and IO (red) in neutrino masses and for $\tan\beta=50$.}
\label{fig1}
\end{figure}

\begin{figure}[!ht]
\centering
\subfigure{\includegraphics[width=0.42\textwidth]{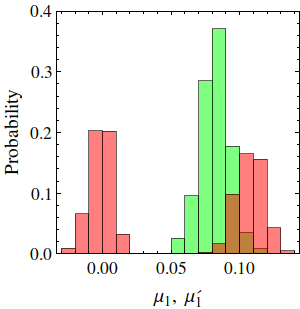}}\quad
\subfigure{\includegraphics[width=0.42\textwidth]{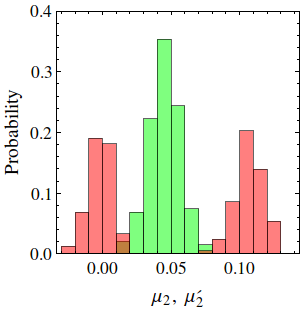}}\\
\subfigure{\includegraphics[width=0.42\textwidth]{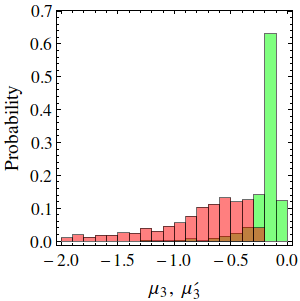}}\quad
\subfigure{\includegraphics[width=0.46\textwidth]{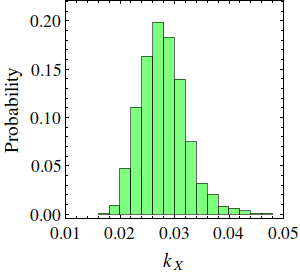}}\\
\caption{The distributions of bulk mass parameters fitted with $\chi^2_{\rm min}/\nu < 3.27$ (or $p >0.001$) in case
of NO and $\tan\beta=50$. The green (red) distribution corresponds to unprimed (primed) bulk mass parameters.}
\label{fig2}
\end{figure}

\begin{figure}[!ht]
\centering
\subfigure{\includegraphics[width=0.42\textwidth]{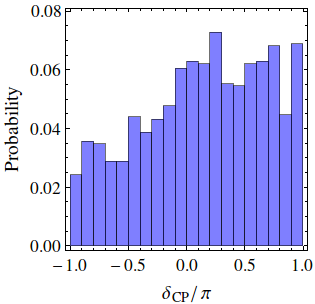}}\quad
\subfigure{\includegraphics[width=0.4\textwidth]{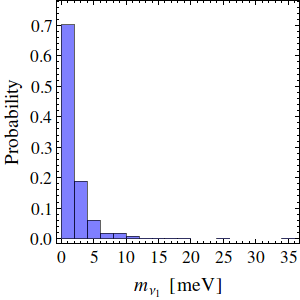}}\\
\subfigure{\includegraphics[width=0.42\textwidth]{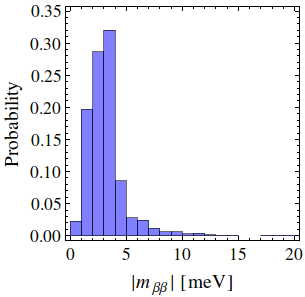}}\quad
\subfigure{\includegraphics[width=0.43\textwidth]{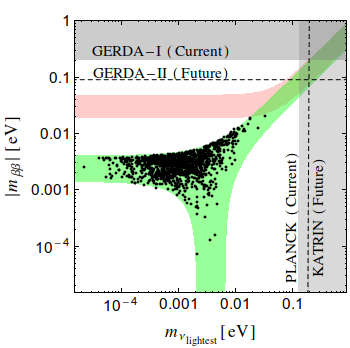}}
\caption{The Yukawa interactions of PS brane: prediction for various observables obtained for $p>0.001$ (corresponding to $\chi^2_{\rm min}/\nu < 3.27$ for $\nu=8$) in case of normally ordered neutrino masses and $\tan\beta=50$. The black points in the bottom-right panel are model predictions while the green (red) regions are the allowed ranges for $|m_{\beta \beta}|$ and the lightest neutrino mass in case of NO (IO). The different horizontal and vertical grey bands correspond to the currently excluded regions by GERDA-I \cite{Agostini:2013mzu} and Planck Cosmic Microwave Background measurements and galaxy clustering information from the Baryon Oscillation Spectroscopic Survey \cite{Giusarma:2013pmn}. The dashed lines indicate the near future reach of GERDA-II and KATRIN \cite{Osipowicz:2001sq} experiments. }
\label{fig3}
\end{figure}

We now discuss the second kind of approach in which we do not fit the fundamental Yukawa couplings of the theory. We treat them as free ${\cal O}(1)$ parameters and restrict their absolute values within the range 0.5 - 1.5, allowing arbitrary phases. For given values of these couplings, we minimize the $\chi^2$ function with respect to the bulk masses and Higgs mixing angles. We repeat this procedure many times, each time generating randomly a new set of Yukawa couplings. We fit $17$ observables with respect to $9$ free parameters (7 bulk masses and 2 Higgs  mixing angles), leaving $\nu=8$ degrees of freedom (dof). The analysis is performed for $\tan\beta=50$ and for NO and IO in the neutrino masses. The results are displayed in Fig. \ref{fig1} where we plot the normalized distribution of the minimum $\chi^2/\nu$. One can see a clear preference for the NO with respect to the IO. Even though one obtains a good best fit for IO case in Table \ref{tab3}, this analysis shows that the solution requires more fine-tuning in the underlying Yukawas compared to the one obtained for NO. 
The $\chi^2$ thresholds corresponding to a given probability value $p$ and the number of cases satisfying the thresholds for different $p-$values are listed in Table \ref{tab4}. 
\begin{table}[!ht]
\begin{small}
\begin{center}
\begin{tabular}{lcccc}
 \hline
 \hline
~~~$p-$value & ~~~0.10~~~& ~~~0.05~~~ & ~~~0.01~~~& ~~~0.001~~~\\
$\chi^2_{\rm min}$ (for $\nu=8$)~~~~~ & $\le$ 13.36 & $\le$ 15.51 & $\le$20.09 & $\le$ 26.12\\
\hline
successful cases (NO) & 0.03\% & 0.05\% & 0.15\% & 0.48\%\\
successful cases (IO) & ~~$<10^{-3}$\%~~ & ~~$<10^{-3}$\%~~& ~~$<10^{-3}$\%~~ &
0.005\%\\
\hline
\hline
\end{tabular}
\end{center}
\end{small}
\caption{The rate of successful events obtained for different $p-$values from random samples of
${\cal O}(1)$ Yukawa couplings in case of normal and inverted ordering in the neutrino masses.}
\label{tab4}
\end{table}
For $p \ge 0.001$, we find $0.5 \%$ cases providing the acceptable values of the $\chi^2_{\rm min} \le 26.12$.  The distributions of the bulk mass parameters and physical predictions for the NO case with $p > 0.001$ are given in Fig. \ref{fig2} and \ref{fig3} respectively.  

One finds preference for positive bulk masses for the first and second generations, which are localized close to the $y=0$ brane. The third generation is localized on the PS brane with a negative bulk mass.  From the distributions shown in Fig. \ref{fig2}, it is clear that the SO(10) breaking by $Z_2'$, which distinguishes $\mu_i$ and $\mu'_i$, is crucial in generating realistic fermion masses in this model. This is particularly true for the first two generations where difference between $\mu_i$ and $\mu'_i$ is significant. Notice that this difference is the only source of breaking of the mass degeneracy between the charged leptons and down-type quarks in this model. The $k_X$ parameter is required to be positive and of the order of the bulk masses of the first two generations. Among the observable quantities in the lepton sector, the lightest neutrino mass is predicted to be below 10 meV corresponding  to strongly hierarchical neutrinos. The effective mass of the neutrinoless double beta decay $|m_{\beta\beta}|$ lies in the range 1-5  meV, which is beyond the reach of the current generation of experiments. Future detection of neutrino masses well above $0.05$ eV and/or of $|m_{\beta\beta}|$ well above the range 1-5  meV would rule out the present model. Since the CP violation is coming from anarchic ${\cal O}(1)$ Yukawas, we get no particular preference for the Dirac CP phase in the lepton sector.  The model do not favour specific values also for the Majorana CP phases as revealed from the correlations between the $|m_{\beta \beta}|$ and the lightest neutrino mass in the bottom-right panel in Fig. \ref{fig3}.

\begin{figure}[!ht]
\centering
\subfigure{\includegraphics[width=0.45\textwidth]{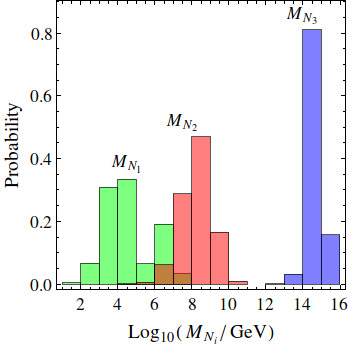}}\quad
\subfigure{\includegraphics[width=0.46\textwidth]{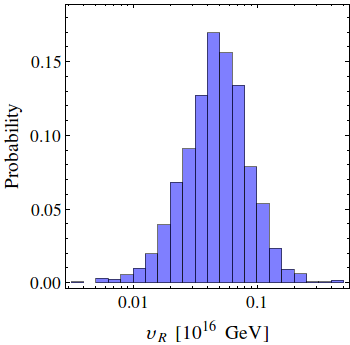}}\quad
\caption{The Yukawa interactions of PS brane: prediction for the masses of RH neutrinos and $v_R = \langle \overline{\Sigma} \rangle^2/\Lambda$ obtained for $p>0.001$ (corresponding to $\chi^2_{\rm min}/\nu < 3.27$ for $\nu=8$) in case of normally ordered neutrino masses and $\tan\beta=50$.}
\label{fig4}
\end{figure}

Since the RH neutrinos are accommodated in $\16$-plets, their masses are predicted once the masses and mixing angles of remaining fermions are fitted. The predictions are displayed in Fig. \ref{fig4}.  
The spectrum of RH neutrinos turns out to be very hierarchical. This is a consequence of the large U(1)$_{\rm X}$ charge of RH neutrinos which generates very large corrections in the bulk masses of the first and second generations making $N_{1,2}$ more sharply localized on $y=0$ brane compared to the other fermions. Since $k_X \ll |\mu'_3|$, the third generation RH neutrino remains localized on the PS brane and one gets $M_{N_3} \approx v_R = \langle \overline{\Sigma} \rangle^2/\Lambda$. We obtain  relatively light spectrum for the first two generation RH neutrinos corresponding to  $M_{N_{2}} \in [10^7,10^{10}]$ GeV and $M_{N_{1}} \in [10^3,10^5]$ GeV. This is in contrasts to generic 4D SO(10) GUT models \cite{Joshipura:2011nn,Dueck:2013gca} where they turn out to be relatively heavier. We also obtain the prediction for $v_R$ after correctly fixing the scale of solar and atmospheric neutrinos. This is shown in  Fig. \ref{fig4}. One finds $\langle \overline{\Sigma} \rangle \approx M_{\rm GUT}$ from the preferred values of $v_R$ which is of the same order as required  by the cancellation of the D-term in Eq. (\ref{DtermPS}). Note that $|\langle \overline{\Sigma} \rangle | > |\langle \Sigma \rangle| \sim M_{\rm GUT}$ is required since $k_X = \sqrt{2} g_5 \langle\45_\Phi\rangle/ \Lambda$ is positive. 

The spectrum of RH neutrinos is strongly hierarchical in our model. In the standard thermal leptogenesis \cite{Fukugita:1986hr} scenario, the final lepton asymmetry is dominated by the lepton number violating decays of the lightest RH neutrino. In this case the successful leptogenesis generically requires \cite{Davidson:2002qv,Buchmuller:2002rq,Blanchet:2006be}
\be \label{N1-bound}
 M_{N_1} \ge 3 \times 10^9~{\rm GeV}~ . \ee
Clearly, this condition is not respected in our model. To further assess the viability of this scenario, we perform a global fit imposing Eq. (\ref{N1-bound}) in our model. We get $\chi^2_{\rm min} \sim 150$ ruling out strongly the possibility of the $N_1$-dominated leptogenesis. An alternative is to consider $N_2$ or $N_3$-dominated leptogenesis, where the lepton flavour effects play an important role \cite{Vives:2005ra}. In this case, the lepton asymmetry is mainly generated by $N_2$ or $N_3$ decays. The lepton doublets produced in such decays get completely incoherent in flavour space before the wash-out by the light RH neutrinos becomes active \cite{Abada:2006fw,Abada:2006ea,Nardi:2006fx,Engelhard:2006yg,Antusch:2010ms}. The wash-out acts individually on each flavour asymmetry and it is less efficient. In this case a certain combination of flavour asymmetry remains protected from the light RH neutrinos wash-out \cite{Vives:2005ra}. We have checked this possibility in our model using the best fit solution reported in Table \ref{tab3} and in the Appendix. We find that $N_2$ is too light to create a sufficient asymmetry, while most of the asymmetry generated by $N_3$ is eventually washed out by $N_2$ and $N_1$, since these particles have sufficiently large couplings with lepton doublets and Higgs. Therefore, our preliminary investigations performed on the best fit solution indicate that leptogenesis cannot be successfully realized in this model. However a detailed analysis of this issue performing a global $\chi^2$ fit including the constraints imposed by flavoured leptogenesis would be required before ruling out leptogenesis in our model, which goes beyond the scope of the present work.

\subsection{Results for the SO(10) brane}
We now investigate the naturalness of anarchic Yukawas on the SO(10) brane, as briefly discussed at the end of sections 2 and 3. The fermion mass relations are similar to the one already derived for the modified KL model in \cite{Feruglio:2014jla}. With respect to the modified KL model, we have three more bulk masses and several new Yukawa couplings in this model. We obtain good global fits for both NO and IO, when $\tan\beta=50$. Therefore we perform the second type of analysis in which we fit the 7 bulk mass parameters and 8 Higgs mixing parameters (see \cite{Feruglio:2014jla} for the details), by taking a flat random distribution for all the ${\cal O}(1)$ anarchical parameters. The ranges of these parameters is chosen as in the previous case. Because of the new parameters coming from the Higgs mixing, with respect to the PS brane, we now have only $\nu=2$ degrees of freedom. 

\begin{figure}[!ht]
\centering
\subfigure{\includegraphics[width=0.6\textwidth]{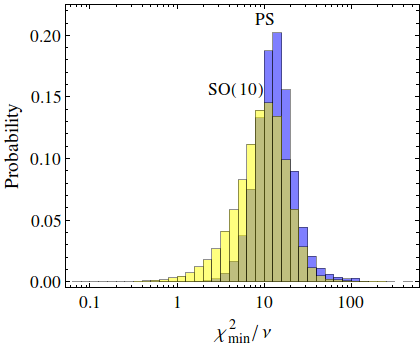}}
\caption{A comparison between the Yukawa interactions on PS ($y=\pi R/2$) and SO(10) ($y=0$) branes. The distributions are obtained for the normal ordering in the neutrino masses and for $\tan\beta = 50$.}
\label{fig5}
\end{figure}

To compare this case to the previous one, we plot the distributions of $\chi^2/\nu$ for both of them and for NO in neutrino masses in Fig. \ref{fig5}.
As it can be seen, both the distributions peak around similar values of $\chi^2/\nu$. The SO(10) case however has a relatively broader distribution leading to more successful cases for a given $p$-value. We get 7\%, 15\% and 30\% successful cases for $p$-values greater than 0.05, 0.01 and  0.001 respectively (the corresponding thresholds for $\chi^2_{\rm min}$ for $\nu=2$ dof are 5.99, 9.21 and 13.82). The substantial increase in the success rate in this case compared to that with Yukawas on the PS brane is attributed to the fact that we have six more mixing parameters providing more freedom in fitting the fermion masses and mixing angles starting from random Yukawa couplings. A similar improvements can be seen by comparing the success rates of this case with those of the modified KL model in \cite{Feruglio:2014jla}. The improved success rates in this case is due to three more bulk mass parameters, which allows better fitting of the data.

\begin{figure}[!ht]
\centering
\subfigure{\includegraphics[width=0.42\textwidth]{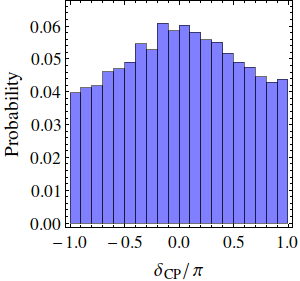}}\quad
\subfigure{\includegraphics[width=0.4\textwidth]{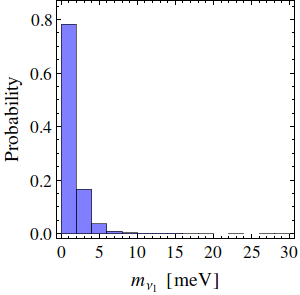}}\\
\subfigure{\includegraphics[width=0.42\textwidth]{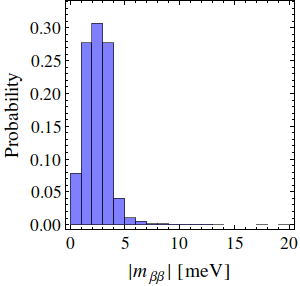}}\quad
\subfigure{\includegraphics[width=0.43\textwidth]{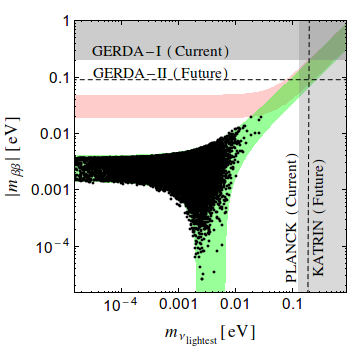}}
\caption{The Yukawa interactions on SO(10) brane: prediction for various observables obtained for the successful cases corresponding to $p > 0.001$ (or $\chi^2_{\rm min}/\nu < 6.91$ for $\nu=2$) in case of normally ordered neutrino masses and $\tan\beta=50$. See Fig. \ref{fig3} for detailed description.}
\label{fig6}
\end{figure}

\begin{figure}[!ht]
\centering
\subfigure{\includegraphics[width=0.45\textwidth]{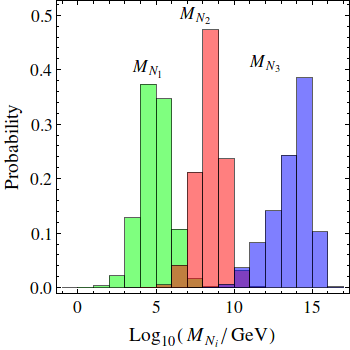}}\quad
\subfigure{\includegraphics[width=0.47\textwidth]{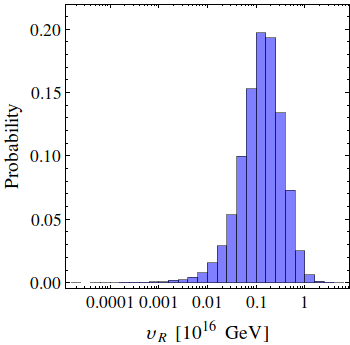}}\quad
\caption{The Yukawa interactions on SO(10) brane: prediction for  for the masses of RH neutrinos and $v_R = \langle \overline{126}_H \rangle $ obtained for the successful cases corresponding to $p > 0.001$ (or $\chi^2_{\rm min}/\nu < 6.91$ for $\nu=2$) in case of normally ordered neutrino masses and $\tan\beta=50$.}
\label{fig7}
\end{figure}

The predictions for the various observables in the successful cases, corresponding to the $p \ge 0.001$, are displayed in Figs. \ref{fig6} and \ref{fig7}.
All the predictions are very similar to those obtained in the case of Yukawas on the PS brane and modified KL model in \cite{Feruglio:2014jla}. This shows that these predictions depend almost entirely on the 
dynamics of the bulk that, generating different zero-mode profiles, distinguishes the various fermion sectors. On the contrary, details of the brane interactions affects only very mildly our results.
The main difference arising from the brane interactions in the different cases is the number of free ${\cal O}(1)$ parameters and Higgs mixing parameters. Our study shows that
when the number of bulk mass parameters and Higgs mixing parameters increases also the rate of success, normalized to the number of degrees of freedom, increases.

\section{Conclusion and discussion}
\label{conclusion}
Grand unified theories, proposed more than forty years ago, provide an elegant synthesis of electroweak and strong interactions, which greatly clarifies some of the crucial aspects of the SM such as
particle classification, gauge anomaly cancellation, quantization of the electric charge and diversification of the gauge coupling constants. In SO(10) grand unified theories
one fermion generation fits in a single representation of the gauge group, leaving room for a right-handed neutrino, which naturally gives rise to neutrino masses through the see-saw mechanism.
This impressive feature is at the heart of the well-known problem of finding an acceptable description of quark and lepton masses and mixing angles, which, in the low-energy data,
do not reflect at all such a complete particle unification. While it is certainly possible to accommodate the observed fermion spectrum by exploiting the most general Yukawa interactions
allowed by the theory, not much is gained with respect to the SM since a huge hierarchy in the Yukawa coupling is needed to reproduce the data.
             
An attractive framework where all the fundamental Yukawa couplings are of order one can be realized, even in SO(10) grand unified theories, through the localization of the profiles for
the zero-mode fermions in an extra dimension. Yukawa interactions are defined on one brane and the hierarchy among fermion masses of different generations
depends exponentially on the bulk fermion masses. Quarks and leptons can be further differentiated by inducing a breaking of the SO(10) symmetry in the bulk mass parameters. 
While the generic ingredients of this construction are well-defined, a considerable freedom is left in model building, depending on the specific implementation of the idea.
In a previous work we relied on a spontaneous breaking of the grand unified symmetry, at the cost of introducing large SO(10) representations for the symmetry breaking sector
with a non-trivial mechanism to solve the doublet-triplet splitting problem. In the present work we have fully exploited the capabilities of the higher-dimensional construction, which
allows for gauge symmetry breaking through compactification and offers a more economic solution to the doublet-triplet splitting problem. Since, compared to our previous model, the new construction significantly
alters the allowed bulk masses and the Yukawa interactions, we think it deserves an accurate study of its properties, to assess whether
the description of fermion masses and mixing angles remains the same or it undergoes  major modifications.
 
We propose a supersymmetric SO(10) model formulated in five dimension. The extra dimension is compactified on an orbifold $S^1/(Z_2\times Z_2')$ and plays a key role in breaking the symmetries of the model.
The compactification breaks N=2 SUSY down to N=1 SUSY in 4D and, at the same time, breaks SO(10) down to the Pati-Salam group SU(4)$_{\rm C} \times$SU(2)$_{\rm L} \times$SU(2)$_{\rm R}$.
A further reduction of the gauge symmetry is realized spontaneously, through a symmetry breaking sector including an SO(10) adjoint, automatically present in this 5D construction,
and additional brane multiplets included with the purpose of canceling the D-terms of the theory.
Below the GUT scale the residual gauge symmetry is that of the SM, which can be finally broken down to SU(3)$_{\rm C}\times$U(1)$_{\rm em}$ by
a set of electroweak doublets localized on the PS brane.  Matter multiplets, introduced in ${\bf 16}$ representations of the GUT group as bulk fields, develop profiles for the zero-modes that are 
localized in specific regions of the extra dimensions. A different localization for the zero-mode profiles of the SM fermions is achieved by different bulk masses. 
As in the original Kitano-Li model, a universal parameter, proportional to the VEV of the adjoint of SO(10), allows to distinguish the different SU(5) components inside a ${\bf 16}$ representation. Moreover our framework allows for independent bulk masses for electroweak singlets and doublets of the various generations. Yukawa interactions can be localized either on the SO(10) or on the PS brane.
While we briefly commented on the first possibility, in our study we mainly concentrated on the PS case, since it offers the possibility of introducing an economic Higgs sector,
which in particular automatically solves the DT splitting problem.
         
Our model, with Yukawa interactions on the PS branes, has seven parameters controlling the bulk masses and two Higgs-mixing parameters, plus a large number of ${\cal O}(1)$ Yukawa couplings.
By fitting an idealized set of data, extrapolated at the GUT scale from the observed fermion masses and mixing angles, we find that the agreement is not trivial and requires a large value of  $\tan\beta$.                 
Moreover the case of inverted ordering in the neutrino mass spectrum requires much more fine-tuning in the Yukawa couplings than the case of normal ordering. The lightest neutrino mass is predicted to be below 10 meV and the effective mass of the neutrinoless double beta decay $|m_{\beta\beta}|$ lies in the range 1-5  meV. The model can be falsified by the observation of either a non vanishing neutrino mass at KATRIN \cite{Osipowicz:2001sq} or $|m_{\beta\beta}|$ at the next generation of experiments. We find no preference for the Dirac CP phase of the lepton sector and the spectrum of RH neutrinos is predicted to be very hierarchical, which unfortunately is incompatible with the generation of the observed baryon asymmetry through thermal leptogenesis.
         
It is remarkable that all these features remain essentially unchanged in several versions of the SUSY SO(10) model in 5D, having in common the property of describing the fermion spectrum
through a set of zero-mode profiles able to distinguish the three generations and the different SU(5) components inside a ${\bf 16}$ representation. All the remaining features of the model such as
the number of independent Yukawa couplings on the branes, the number of Higgs mixing parameters, the additional possibility of distinguishing weak doublets and singlets through the bulk masses,
seem to play a secondary role which, at most, can influence the success rate of the model when statistical tests are performed. We conclude that the results are rather robust against
modifications of the basic framework. We find very interesting the possibility of combining in a realistic scheme the anarchy of the underlying Yukawa couplings with the unification of one fermion generation implied by the
SO(10) GUT. On the weak side, as all models based on a large number of independent ${\cal O}(1)$ parameters, it is not possible to plan precision tests of these ideas to fully exploit the accuracy
of existing data.
          

\begin{acknowledgments}
We thank Enrico Nardi for useful correspondence on leptogenesis.
We acknowledge partial support from the European Union FP7 ITN INVISIBLES
(Marie Curie Actions, PITN-GA-2011-289442). K.M.P. thanks the Department of Physics and Astronomy of the University of Padova for its support.
\end{acknowledgments}
 

\appendix
\section{Parameters obtained for the best fit solutions}
We provide the set of input parameters obtained for the best fit solutions corresponding to normal and inverted neutrino mass spectrum and $\tan\beta = 50$ as presented in the Table \ref{tab3}. 

\subsection{Normal ordering}
The values of the Yukawa matrices and bulk masses appearing in Eqs. (\ref{4dyukawa}, \ref{massmatrices}, \ref{seesaw}) at $\chi^2_{\rm min}\approx 0$ are as the following. We have removed some unphysical phases by redefining the fields.
\beqa 
Y_{u}&=&
\left(
\begin{array}{ccc}
  0.55863 ~e^{  -0.49590i} & 0.94275 & 1.23911 ~e^{ -1.10433i}\\
 0.74927 &1.49374 & 0.66883 \\
0.50804 ~e^{  0.22131i}& 0.50000 &  1.26156 ~e^{  -0.86038i}\\
\end{array}
\right)
~, \nonumber \\
Y_{d}&=&\left(
\begin{array}{ccc}
  0.64691 ~e^{  -0.51014i} & 0.71998 ~e^{ -0.81349i} & 0.52244 ~e^{  2.72841i} \\
 0.80610 ~e^{  1.57886i} & 0.57351 ~e^{  0.23467i} &  0.50398 ~e^{ 0.47936i} \\
 1.01632 ~e^{  -0.85648i} & 0.59252 ~e^{ -1.77531i} &  0.63639 ~e^{ -2.92490 i} \\
\end{array}
\right)
~, \nonumber \\
Y_{R}&=&\left(
\begin{array}{ccc}
1.10716~ e^{ 0.17875 i} & 0.70519 ~e^{  0.94555i} &  0.81595 ~e^{ -0.75271i} \\
0.70519 e^{  0.94555i} &  1.30773 ~e^{   2.93543i} & 1.07719 ~e^{  -0.17411i} \\
0.81595 e^{   -0.75271i} & 1.07719 ~e^{  -0.17411i} &  0.71443 ~e^{ 1.37417i}  \\
\end{array}
\right)~. 
\eeqa
The corresponding bulk mass parameters are:
\beqa
\{\mu_1,~\mu_2,~\mu_3\} &=& \{0.049590,\,0.020895,\,-0.139245\}~,\nonumber  \\
\{\mu_1^\prime,~\mu_2^\prime,~\mu_3^\prime\} &=& \{0.066244,\,-0.013373,\,-0.463361\}~, \\
k_X&=& 0.042394~.\nonumber 
\eeqa

From the above parameters the profile matrices in Eq. (\ref{Fmat}) for various SM fermions can be expressed in terms of powers of the Cabibbo angle $\lambda$ as below.
\beqa
F_{Q}=\lambda ^{0.6} \left(
\begin{array}{ccc}
 \lambda ^{3.1} & 0 & 0 \\
 0 & \lambda  ^{2.3}  & 0 \\
 0 & 0 & 1 \\
\end{array}
\right) \!\!&,&\!\!~
F_{d^c}=  \frac{1}{\lambda ^{0.1}} \left(
\begin{array}{ccc}
 \lambda ^{0.8} & 0 & 0 \\
 0 & \lambda ^{0.5} & 0 \\
 0 & 0 & 1 \\
\end{array}
\right), \nonumber \\
F_{L}=\lambda ^{0.2} \left(
\begin{array}{ccc}
 \lambda ^{0.4} & 0 & 0 \\
 0 & \lambda ^{0.3} & 0 \\
 0 & 0 & 1 \\
\end{array}
\right) \!\!&,&\!\!~
F_{N^c}=\lambda ^{0.2} \left(
\begin{array}{ccc}
 \lambda ^{9.4} & 0 & 0 \\
 0 & \lambda ^{6.8} & 0 \\
 0 & 0 & 1 \\
\end{array}
\right),~\nonumber \\
F_{u^c} = F_{e^c} \!\!&=&\!\!\lambda ^{0.1} \left(
\begin{array}{ccc}
 \lambda ^{4.2} & 0 & 0 \\
 0 & \lambda ^{1.9} & 0 \\
 0 & 0 & 1 \\
\end{array}
\right).
 \eeqa

\subsection{Inverted ordering}
The values of the Yukawa matrices and bulk masses appearing in Eqs. (\ref{4dyukawa}, \ref{massmatrices}, \ref{seesaw}) at $\chi^2_{\rm min}\approx 0.96$ are as the following. We have removed some unphysical phases by redefining the fields.
\beqa 
Y_{u}&=&
\left(
\begin{array}{ccc}
 1.05063 ~e^{ -2.27438i} & 0.50197 &  0.50108 ~e^{ 0.65794i} \\
 1.28888 & 0.95572 &  0.95749 \\
 1.32079 ~e^{  1.96363i} & 0.84379 &  1.03615 ~e^{-1.69586i}  \\
\end{array}
\right)
~, \nonumber \\
Y_{d}&=&\left(
\begin{array}{ccc}
0.51388~e^{ -2.46719i} &  0.50192~e^{ 1.08880i} & 0.72278~e^{ 1.00274i} \\
1.47850~e^{ -1.34548i} & 0.63988~e^{ 1.91581i} &  0.62270~e^{ 0.06790i} \\
0.68440~e^{ -1.92037i} & 0.52781~e^{ 1.82283i} &  0.50618~e^{ 1.03128i} \\
\end{array}
\right) 
~, \nonumber \\
Y_{R}&=&\left(
\begin{array}{ccc}
 1.32057~e^{ -1.64402i} & 1.34754~e^{-2.56275i} & 0.62345~e^{ 1.12638i} \\
1.34754~e^{ -2.56275i} & 1.44530~e^{1.87202i} &  0.57696~e^{  -0.04777i} \\
 0.62345~e^{  1.12638i} & 0.57696~e^{ -0.04777i} &  0.62830~e^{ 2.31181i} \\
\end{array}
\right) ~. 
\eeqa
The corresponding bulk mass parameters are:
\beqa
\{\mu_1,~\mu_2,~\mu_3\} &=& \{0.056934,\,0.023583,\,-0.212866\}~,\nonumber  \\
\{\mu_1^\prime,~\mu_2^\prime,~\mu_3^\prime\} &=& \{0.088673,\,-0.025229,\,-0.421995\}~, \\
k_X&=& 0.045419~. \nonumber 
\eeqa

From the above parameters the profile matrices in Eq. (\ref{Fmat}) for various SM fermions can be expressed in terms of powers of the Cabibbo angle $\lambda$ as below.
\beqa
F_{Q}=\lambda ^{0.4} \left(
\begin{array}{ccc}
 \lambda ^{3.6} & 0 & 0 \\
 0 & \lambda ^{2.6} & 0 \\
 0 & 0 & 1 \\
\end{array}
\right) \!\!&,&\!\!~
F_{d^c}=  \frac{1}{\lambda ^{0.04}} \left(
\begin{array}{ccc}
 \lambda ^{0.8} & 0 & 0 \\
 0 & \lambda ^{0.4} & 0 \\
 0 & 0 & 1 \\
\end{array}
\right), \nonumber \\
F_{L}=\lambda ^{0.1} \left(
\begin{array}{ccc}
 \lambda ^{0.5} & 0 & 0 \\
 0 & \lambda ^{0.4} & 0 \\
 0 & 0 & 1 \\
\end{array}
\right) \!\!&,&\!\!~
F_{N^c}=\lambda ^{0.3} \left(
\begin{array}{ccc}
 \lambda ^{10.6} & 0 & 0 \\
 0 & \lambda  ^{6.9} & 0 \\
 0 & 0 & 1 \\
\end{array}
\right),~\nonumber \\
 F_{u^c} = F_{e^c} \!\!&=&\!\! \lambda ^{0.1} \left(
\begin{array}{ccc}
 \lambda ^{4.9} & 0 & 0 \\
 0 & \lambda ^{1.6} & 0 \\
 0 & 0 & 1 \\
\end{array}
\right)~~~.
 \eeqa

\bibliographystyle{apsrev4-1}
\bibliography{ref-so10.bib}
\end{document}